\documentclass[aps,prl,showpacs,floats,twocolumn,floats,superscriptaddress,floatfix]{revtex4-1}
\usepackage{amssymb}
\usepackage{graphicx}
\usepackage{bm}
\usepackage{amsfonts}
\usepackage{color}
\usepackage{amsmath}    
\usepackage{epsfig}
\usepackage{hyperref}   
\usepackage{mathrsfs}

\usepackage{amssymb}
\usepackage{graphicx}
\usepackage{bm}
\usepackage{amsfonts}
\usepackage{color}
\usepackage{amsmath}    
\usepackage{epsfig}

\newcommand{\kg}{{K_{\scriptscriptstyle \mathrm{G}}}}

\newcommand{\pkg}{\mathrm{P}_{\!\!\! _{K_{\scriptscriptstyle
        \mathrm{G}}}}}

\newcommand{\ue}{\mathrm{e}}
\newcommand{\ui}{\mathrm{i}}

\newcommand{\ictsaddress}{International Centre for
  Theoretical Sciences, Tata Institute of Fundamental Research,
  Bangalore 560089, India}
\newcommand{\OCA}{Universit\'e C\^ote d’Azur, CNRS, OCA, Laboratoire J.-L. Lagrange, Nice, France}

\newcommand{\UCA}{Universit\'e C\^ote d’Azur, CNRS, Institut de Physique de Nice, Nice, France}
\begin{document}
\title{Suppressing thermalisation and constructing weak solutions in truncated inviscid equations of hydrodynamics: Lessons from the Burgers equation}
\author{Sugan D. Murugan}
\email{sugan.murugan@icts.res.in}
\affiliation{\ictsaddress}
\author{Uriel Frisch}
\email{uriel@oca.eu}
\affiliation{\OCA}
\author{Sergey Nazarenko}
\email{snazar1@yahoo.co.uk}
\affiliation{\UCA}
\author{Nicolas Besse}
\email{Nicolas.Besse@oca.eu}
\affiliation{\OCA}
\author{Samriddhi Sankar Ray}
\email{samriddhisankarray@gmail.com}
\affiliation{\ictsaddress}
\keywords{Thermalisation $|$ Weak Solutions $|$ Finite-time Blow-up $|$ Euler and Burgers equations}
\begin{abstract}
Finite-dimensional, inviscid equations of hydrodynamics, obtained through a Fourier-Galerkin projection,
thermalise with an energy equipartition. Hence, numerical solutions of such inviscid equations, which typically
\textit{have} to be Galerkin-truncated, show a behaviour at odds with the
parent equation. An important consequence of this is an uncertainty in the measurement of the temporal
evolution of the distance of the complex singularity from the real domain leading to a
lack of a firm conjecture on the finite-time blow-up problem in the
incompressible, three-dimensional Euler equation.  We now propose, by using the
one-dimensional Burgers equation as a testing ground, a novel numerical recipe,
named \textit{tyger purging}, to arrest the onset of thermalisation and hence
recover the \textit{true dissipative} solution. Our method, easily adapted for
higher dimensions, provides a tool to not only tackle the celebrated blow-up problem but also to 
obtain weak and dissipative solutions---conjectured by Onsager and numerically
elusive thus far---of the Euler equation.
\end{abstract}
\maketitle

\paragraph*{Introduction:} Non-linear, partial differential equations of hydrodynamics, such as the
inviscid the one-dimensional Burgers or the three-dimensional Euler equations,
are often studied, numerically and theoretically, by projecting them on to a
Fourier subspace with a finite number of modes bounded by a (large) wavenumber
$\kg$. This projection (defined precisely later), known as a
Galerkin-projection, ensures that unlike the parent partial differential
equation (PDE) which has an infinite number of degrees of freedom, the
Galerkin-truncated equation is constrained to have only finitely many Fourier
modes.  Consequently, the resulting finite-dimensional, inviscid equations of
hydrodynamics, such as the three-dimensional (3D) incompressible Euler
equations or the one-dimensional (1D) Burgers equation, conserve both energy
and phase-space, leading to solutions which thermalise in a finite-time. These
solutions are thus completely different from the solutions of the actual
partial differential equation, from which they derive, with infinite degrees of
freedom~\cite{Hopf,Lee}. 

In recent years however, this area has received renewed
interest~\cite{Ray-Review}---spanning studies in turbulence~\cite{Frisch-PRL-2012,Luca-PRL-2015,Luca-EPJE-2016,Buzzicotti-PRE-2016,Buzzicotti-NJP-2017,Ray-PRF-2018}, bottlenecks and
hyperviscosity~\cite{Frisch-PRL-2008,Frisch-PRL-2013,Banerjee-PRE-2014} to problems of
cross-correlators in condensed matter physics~\cite{OTOC}---beginning with the
work of Majda and Timofeyev~\cite{Majda-PNAS} on the thermalisation of the
Galerkin-truncated, 1D inviscid Burgers equation.  Subsequently,
Cichowlas, \textit{et al}.~\cite{Cichowlas-PRL}, through state-of-the-art
direct numerical simulations (DNSs) showed the existence of similar thermalised
states in the Galerkin-truncated 3D incompressible Euler equation (see, also, Ref.~\cite{Krstulovic}). However the
precise mechanism by which solutions thermalise  was discovered later by Ray,
\textit{et al}.~\cite{Ray-PRE11} who showed that thermalisation was triggered
through a resonant-wave-like interaction leading to localised structures
christened \textit{tygers} (see, also, Refs.~\cite{PereiraPRE,BrachetPRF,VenkataramanPRSA}).  

Understanding Galerkin-truncated systems assumes a special importance when
numerically studying inviscid equations for the problem of finite-time blow-up
of the incompressible Euler equation (under suitable conditions). A  way to
conjecture for or against a finite-time singularity is to numerically solve the
Euler equation and measure the width of the analyticity strip
$\delta$~\cite{Sulem1983}, i.e., the distance to the real domain of the
nearest complex singularity.  By assuming analyticity, at least up to a
hypothetical time of blow-up $t_*$, this procedure reduces to measuring the
Fourier modes of the velocity field $\hat{u}_k \sim \exp{[-\delta(t)k]}$
(ignoring vectors for convenience), for large wavenumbers $k$, and thence,
$\delta$ as a function of time $t$. Therefore, a numerically compelling
\textit{proof} for finite-time blow-up is to show $\delta(t) \to 0$ in a
finite time.

Simple as it sounds, such an approach unfortunately runs into a severe problem
in its implementation. To solve such equations on the computer, one has to make
them finite-dimensional through a Galerkin truncation. Solutions to
such truncated equations thermalise, beginning at small scales (or large
wavenumbers $k$) in a finite time. Hence, asymptotically at large wavenumbers
the Fourier modes of the velocity field grow as a power law $\hat{u}_k \sim
k^{d-1}$ (energy equipartition), where $d$ is the spatial dimension, and not
fall-off exponentially from  which the width of the analyticity strip can be
extracted. Hence, the measurement of $\delta(t)$ becomes unreliable
\textit{soon enough} to prevent us from making a reasonable conjecture of if
and when $\delta(t)$ might vanish~\cite{BustamantePRE}. Therefore, in order to
have a more reliable measurement of $\delta(t)$ for times long enough to
conjecture on whether there is a finite-time blow-up of, e.g., the 3D,
incompressible, Euler equation, it is vital to have a (numerical)
prescription---without resorting to viscous damping---which prevents the
solutions from thermalising.

\begin{figure*}
\includegraphics[width=0.49\linewidth]{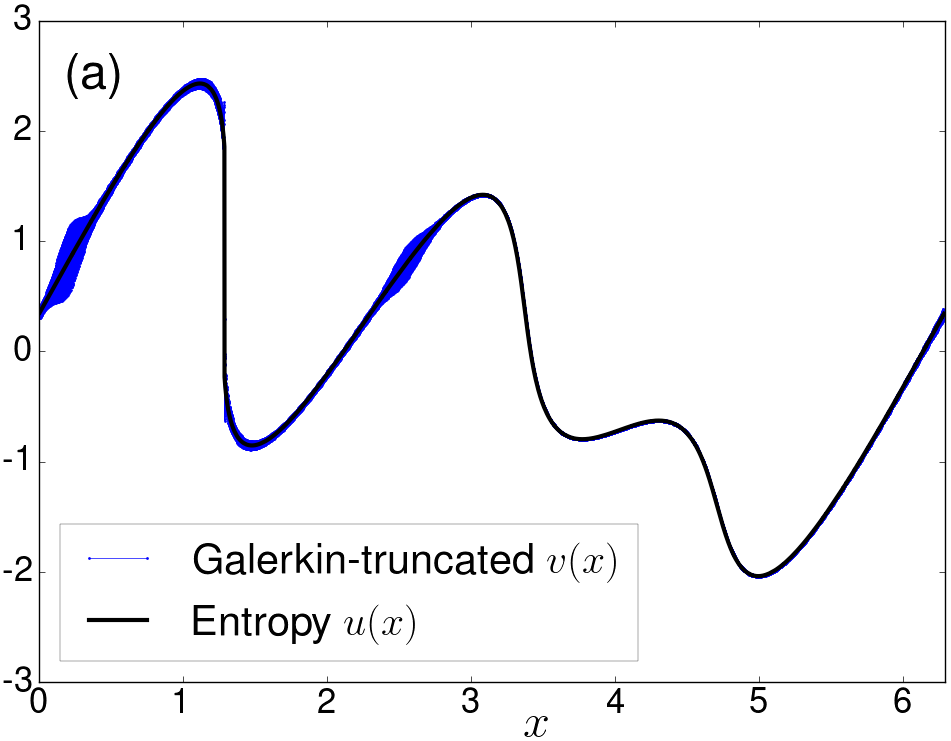}
\includegraphics[width=0.49\linewidth]{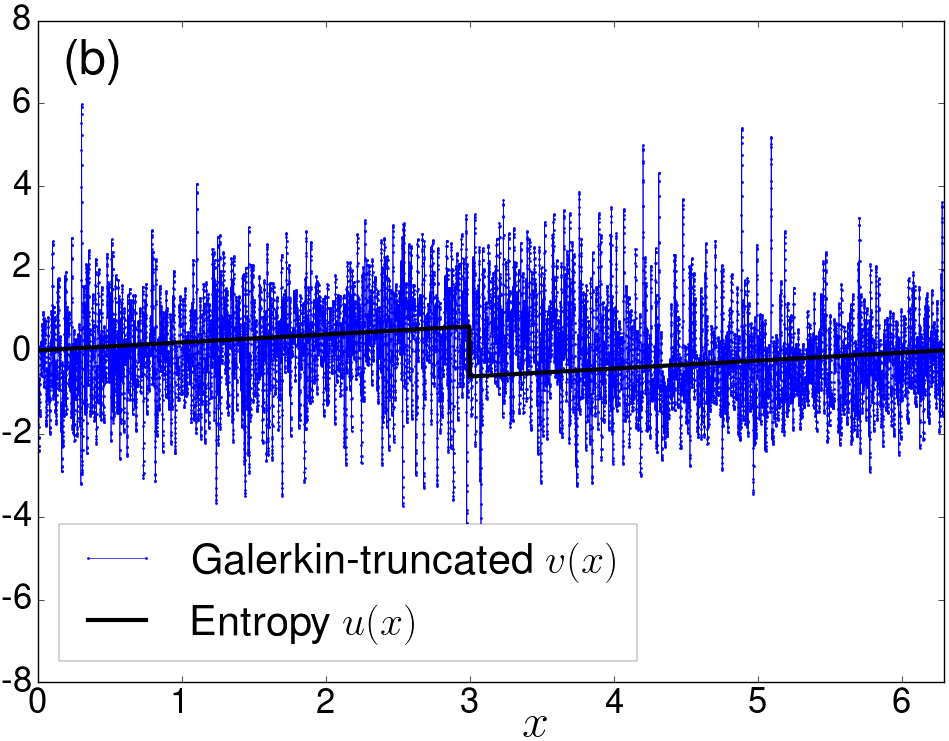}
\caption{Representative plots, for $\kg = 1000$,  of the Galerkin-truncated $v$ (blue) and entropy $u$ (black) solutions of the Burgers 
equation at (a) $t = 0.24 \gtrsim t_* $ and (b) $t  = 5.0 \gg t_*$.  
For the Galerkin-truncated solution, panel (a) shows signatures of impending thermalisation through 
	the birth of \textit{tygers} while panel (b) shows the fully thermalised solutions. (We refer the reader to the youtube 
	link~\url{https://www.youtube.com/watch?v=QiioybbVi6M} for a 
	movie of the time evolution of the Galerkin-truncated equation (and the entropy solution) with a single-mode initial condition for clarity.)}
\label{thermalisation}
\end{figure*}

We now propose such a recipe and show how the Galerkin-truncated equation can be
modified mildly to obtain solutions which do not thermalise. This allows us 
to obtain numerically (a) more reliable estimates of the widths of the
analyticity strip and (b) weak, but dissipative, solutions (henceforth called weak-dissipative, 
for convenience) of inviscid equations.

The reasons which motivates this study are of course fundamentally important
for the 3D Euler equations and less so for the 1D Burgers equation.  However
the process and mechanisms of thermalisation was best understood by resorting
to the 1D Burgers
equation~\cite{Majda-PNAS,Ray-PRE11,Ray-Review,VenkataramanPRSA}; in the same
spirit, we now outline and present results for the efficacy of the tyger
purging method. At the end of this paper, we will return briefly to its
applicability to the problem of the 3D Euler equation as well as contrast our
approach with wavelet-based filtering techniques~\cite{PereiraPRE}. 

\paragraph*{Thermalisation:} Let us begin with the 1D inviscid Burgers equation on a $2\pi$ periodic line
\begin{equation}
\frac{\partial u}{\partial t} + \frac{1}{2}\frac{\partial u^2}{\partial x} = 0
\label{burgers}
\end{equation}
augmented by the initial condition $u_0(x)$ which
is typically a combination of trigonometric functions containing a few Fourier modes. 
Since we work in the space of $2\pi$ periodic solutions, we can 
expand the solution of Eq.\eqref{burgers} in a Fourier series allowing us 
to define the Galerkin projector $\pkg$ as a low-pass filter
which sets all modes with wavenumbers $|k| > \kg$, where $\kg$ is a positive (large) 
integer, to zero via: 
$v(x)=\pkg u(x) =   \sum_{|k| \le \kg} \ue ^{\ui kx}\hat u_k$.

These definitions allow us to write the Galerkin-truncated inviscid Burgers equation 
\begin{equation} 
 \frac{\partial v}{\partial t} + \pkg\frac{1}{2}\frac{\partial v^2}{\partial x} = 0;
\label{gtburgers}
\end{equation}
the initial conditions $v_0 =\pkg u_0$ are similarly projected onto the subspace spanned by $\kg$. 

The solution of the inviscid Burgers equation~\eqref{burgers} show one or more
shocks (determined by $u_0(x)$), after an initial-condition dependent
finite-time $t_*$, through which energy is dissipated for $t>t_*$.
Theoretically, the solution to~\eqref{burgers}, for $t > t_*$, is obtained
by adding a viscous dissipation term $\nu\frac{\partial^2 u}{\partial x^2}$
with $\nu \to 0$ (the inviscid limit), which preserves the finitely many shocks
of the \textit{true} solution.  This generalised solution, in the limit of
vanishing viscosity, converges weakly to the inviscid Burgers equation and is
characterised by a dissipative anomaly: energy dissipation $\epsilon$ remains
finite as $\nu \to 0$. 

In contrast, the Galerkin-truncated equation~\eqref{gtburgers} conserves energy
for all times.  For initial conditions with a finite number of non-vanishing
Fourier harmonic, the solution $v$ mimics rather well that of the inviscid PDE
up to time $t \lesssim t_*$. Indeed, for $t<t_*$, the two solutions are
essentially indistinguishable. However, when the distance of the nearest
(complex) singularity of the un-truncated equation~\eqref{burgers} is within
one Galerkin wavelength ($\sim 2\pi/\kg$) of the real domain (at time $t
\approx t_* - \kg^{-1/3}$), the effect of truncation becomes important. 

For $t > t_*$, the solutions of the truncated-equation and the PDE
are dramatically different: Whereas the former stays smooth, conserves energy,
and start thermalising (beginning at small scales) with an (equipartition) energy spectrum
$\langle|\hat{v}_k|^2\rangle \sim k^0$~\cite{Majda-PNAS}, the latter shows a
monotonic decrease in its kinetic energy (dissipated through the shock(s)) and
an associated scaling $\langle|\hat{u}_k|^2\rangle \sim k^{-2}$. (The angular brackets used
in calculating the energy spectrum denotes suitable ensemble averages.) Thus 
thermalised solutions, inevitable in numerical solutions of the Galerkin-truncated 
inviscid equations, are fundamentally different from---and hence
do not converge to---the un-truncated parent PDE. 

We illustrate this phenomenon in Fig.~\ref{thermalisation} by showing the
solutions  of the Galerkin-truncated equation $v$ (in blue),
with $\kg = 1000$, and the entropy solution $u$ (in black) for
(a) an early time $t = 0.24$ ($\gtrsim t_*$) and (b) at a later time $t = 5.0$
($t \gg t_*$); the details of such numerical simulations are given later. 
As discussed above, even at times very close to $t_* \approx 0.23$
(Fig.~\ref{thermalisation}a), the two solutions show a marked
difference---\textit{tygers}---at points which have the same velocity as the shock (and
a positive fluid velocity gradient). At even later times,
(Fig.~\ref{thermalisation}b) we see clear signatures of thermalisation in the
truncated solution having no resemblance to the entropy solution
which, as a consequence of shocks merging in time, has a saw-tooth structure with a single shock.
We refer the reader to Refs.~\cite{Majda-PNAS, Ray-PRE11,
Ray-Review,PereiraPRE,VenkataramanPRSA,BrachetPRF} for more details and the
theory of this process of thermalisation.

\paragraph*{Tyger purging:} All of this leads us to ask if we can, without resorting to viscous
dissipation, actually suppress thermalisation setting in in such truncated
equations and obtain the entropy solution? The short answer is yes as we now report a novel
approach---\textit{tyger purging}---which, through the selective removal of a
narrow, Fourier space, boundary layer near $\kg$ (see below), at discrete time-intervals,
results in the suppression of thermalisation. 

The equation of motion for the purged solution $w$ is, of course, the same as that of the Galerkin-truncated equation 
(with the truncation wavenumber $\kg$)
\begin{equation} 
 \frac{\partial w}{\partial t} + \pkg\frac{1}{2}\frac{\partial w^2}{\partial x} = 0
\label{purged}
\end{equation}
augmented by an additional constraint imposed at discrete times $t_p = t_* + n \tau$ ($n$ = 0,1,2,3 ...):
\begin{equation}
\hat{w}_k = 0 \quad \forall \quad K_p \le k \le \kg.
\label{constraint}
\end{equation}
We call this truncated equation, along with the additional \textit{purging} constraint,
as simply the purged equation. We note
that without the additional constraint, by definition, the solution
$w$ is the same as $v$ obtained from the truncated
equation and hence if purging is done continuously, 
and not discretely, in time, we would end up solving the Galerkin-truncated 
equation~\eqref{gtburgers} but with a truncation wavenumber $K_p$.

We now make the following \textit{ans\"atze} about the inter-purging time $\tau$ and the purging wavenumber $K_p$:
\begin{equation}
\tau = \kg^{-\alpha}; \quad \quad K_p = \kg - \kg^\beta; 
\label{expo}
\end{equation}
with real, positive exponents $\alpha$ and $\beta$ and the immediate constraint that $\beta < 1$.

Before we engage in a detailed numerical analysis, let us estimate,
heuristically, optimal choices of $\alpha$ and $\beta$ keeping in mind that the
purged solution $w$ must converge to the entropy solution $u$ as $\kg \to
\infty$. 

For $t > t_*$, the entropy solution, unlike the truncated solution, is
dissipative: $\varepsilon \equiv \frac{dE}{dt} < 0$, where $E = \frac{1}{2}\displaystyle \sum_{k =
1}^{\infty}|\hat{u}_k|^2$ is the total energy. Indeed, for times $t \sim t_*$, (when 
tygers are just born), the Galerkin-truncated Burgers equation remains conservative 
by the transfer---and subsequent accumulation---of kinetic energy 
$\propto \kg^{-5/3}$ from the ``shock'' to the tygers~\cite{Ray-PRE11}.

By construction, however, purging allows for a finite energy
loss $\Delta E^{\rm P} \equiv \frac{1}{2}\displaystyle \sum_{k =
K_p}^{\kg}|\hat{w}_k|^2$ at intervals of $\tau$ resulting in a rate of loss of
energy $\varepsilon^{\rm P} \equiv \frac{dE^{\rm P}}{dt} \sim \frac{\Delta E^{\rm P}}{\tau}$, where
$E^{\rm P} = \frac{1}{2}\displaystyle \sum_{k = 1}^{\kg}|\hat{w}_k|^2$ is the
total energy of the purged system.  The choice of $\alpha$ and $\beta$ should
ensure that in the limit $\kg \to \infty$, this rate of energy loss should be
$\kg$-independent and comparable to the rate of energy loss of the entropy solution, i.e., 
$\varepsilon^{\rm P} = \varepsilon$.  

It is hard to estimate $\Delta E^{\rm P}$ theoretically without making suitable
assumptions. Since in between two purges, Eq.~\ref{purged} is identical to
the Galerkin-truncated equation, it is reasonable to assume that at the time of purging the
solution $|\hat{w}_k|$ to be a combination of the one coming from the entropy
solution $\hat{u}_k$ and a contribution from the nascent tyger.  The
latter, which is the deviation of the truncated from the entropy
solution, was shown in Ref.~\cite{Ray-PRE11} to be confined to a narrow
Fourier-space boundary layer close to and up to $\kg$ with a form (ignoring 
$\mathcal{O}(1)$ constants in the prefactors as well as the argument of the exponential)
$\frac{1}{\kg}\mathrm{exp}\left[-\frac{\kg-k}{\kg^{1/3}}\right]$. Keeping
these factors in mind, it is easy to show that  $\varepsilon^{\rm P} \sim -\kg^{\alpha + \beta - 2}$. If we now demand, for convergence, that this
rate be independent of $\kg$, we obtain the constraint $\alpha +
\beta = 2$. 

The constraint derived above is useful but it still allows considerable freedom
in choosing $\alpha$ and $\beta$. However, since in between purgings the solution develops only nascent 
\textit{tygers}, we can estimate $\beta$ independently by asking if 
an optimal choice of $K_p$ (thence, $\beta$) leads to an elimination of the boundary layer (and hence the energy 
content $\delta E^{\rm P}$ of the boundary layer) such that \textit{tygers} are suppressed. In other words, 
since Galerkin-truncation leads to a transfer of energy $\sim \kg^{-5/3}$ from the
shock to the \textit{tygers} resulting in an overall conservation of kinetic energy in
the truncated problem,  a successful purging strategy must constraint $\delta E^{\rm P} \approx
\kg^{-5/3}$ thus precisely eliminating the \textit{tygers} which trigger thermalisation and hence 
leading to dissipative solutions.  By using the functional form for the boundary layer for incipient \textit{tygers}~\cite{Ray-PRE11}, it is 
easy to show that
\begin{eqnarray}
	\delta E^{\rm P} &\equiv & \sum_{k = K_p}^{\kg}|\hat{w}_k - \hat{u}_k|^2  \nonumber \\
			 &\approx & \sum_{k = K_p}^{\kg} \frac{1}{\kg^2}\mathrm{exp}\left[-\frac{\kg-k}{\kg^{1/3}}\right] \nonumber \\
	& \sim & \begin{cases} \kg^{\beta - 2}\:\: \text{for} \:\: \beta < 1/3 \\
  \kg^{-5/3}\:\: \text{for}\:\: \beta > 1/3
  .\end{cases}
\label{DeltaEP}
\end{eqnarray}

Equation~\eqref{DeltaEP} leads to the inevitable conclusion that the optimal choice of 
the purging wavenumber is one where $\beta \in [1/3, 1)$ and the energy loss then is actually 
independent of $\beta$ and exactly the same as that which would have triggered thermalisation in the absence of 
purging as long as $\beta \ge 1/3$. Thus, we obtain an independent (theoretical) bound on $\beta  \in [1/3, 1)$ for 
a successful purging.   

Before we turn to detailed numerical simulations to validate these ideas, we make one final remark. 
In numerical simulations, $\delta t$ is typically set by the resolution $\kg$ such that $\delta t \sim \mathcal{O}(\kg^{-1})$. As we 
have noted before, purging if done too frequently would be akin to solving the Galerkin-truncated Burgers equation with $\kg = K_p$. 
This implies that $\tau/\delta t \gg 1$ which, trivially, leads to $\alpha < 1$. Hence, with these insights for $\alpha$ and $\beta$, we revise 
the constraint, estimated heuristically before, to $\alpha + \beta \lesssim 2$.

\paragraph*{Direct Numerical Simulations:} So how effective is purging in obtaining solutions $w$ which resemble the
entropy solution $u$? We answer this by resorting to extensive and detailed numerical simulation of the purged model~\eqref{purged} 
as well the Galerkin-truncated equation~\eqref{gtburgers} for comparison.

\begin{figure*}
\includegraphics[width=0.49\linewidth]{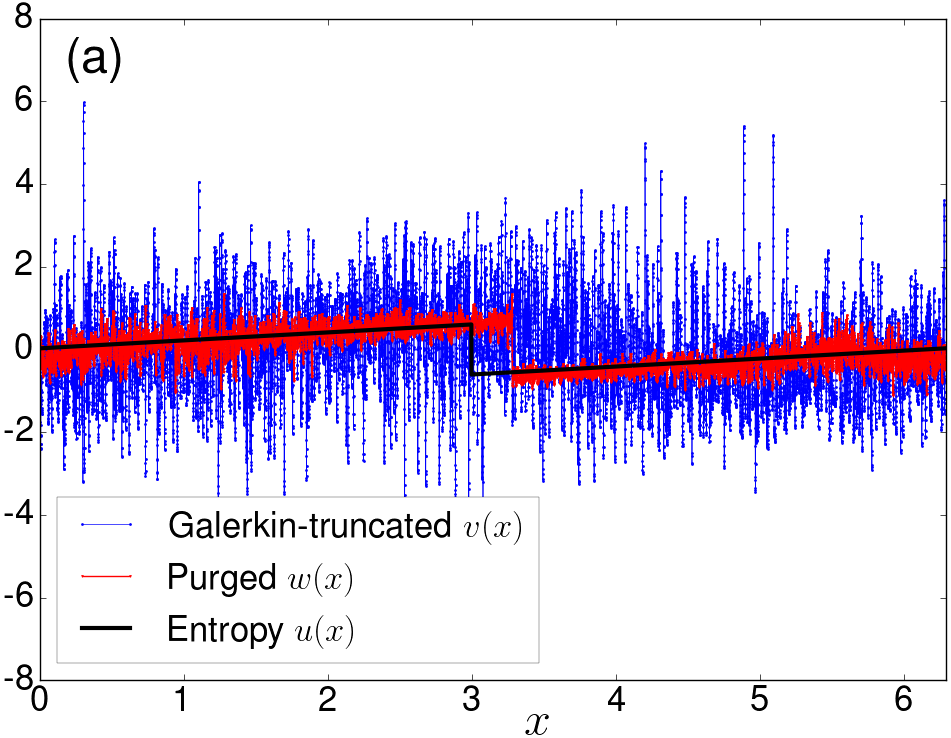}
\includegraphics[width=0.49\linewidth]{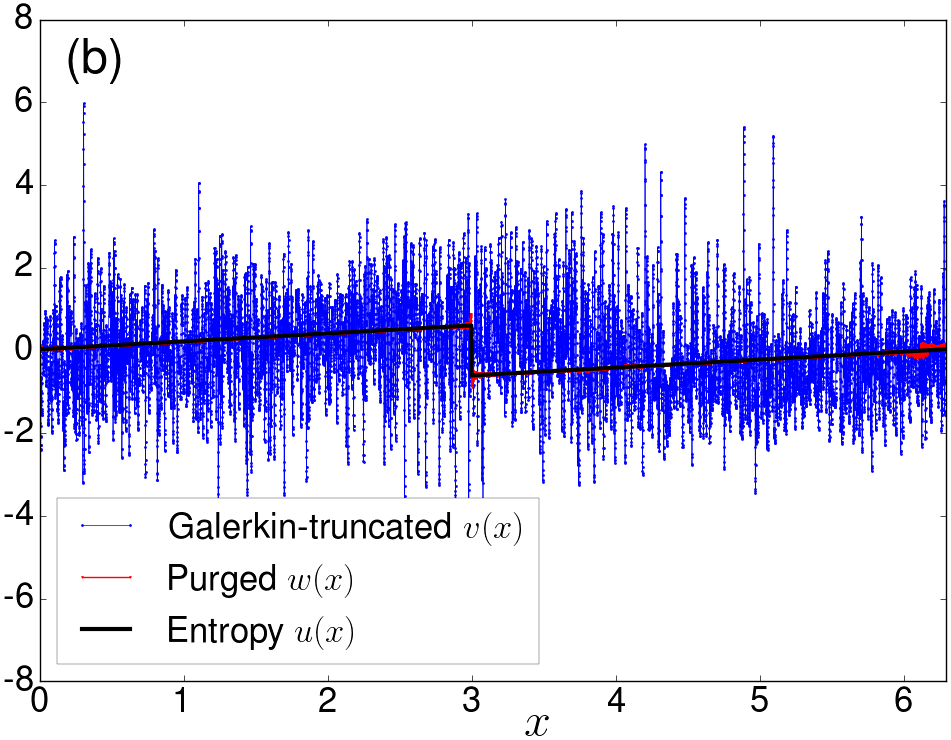}
\caption{Representative plots, for $\kg = 1000$, of the Galerkin-truncated $v$ (blue), the entropy $u$ (black) and the 
purged $w$ (red) solutions of the Burgers 
equation at $t = 5.0$ for (a) $\alpha = 0.6$, $\beta = 0.4$ and (b) $\alpha = \beta = 0.8$. In panel (b), the purged and 
	entropy solutions are quite close to being identical. (We refer the reader to the youtube 
	link~\url{https://www.youtube.com/watch?v=utjyfQUuCIc} for a 
	movie of the full evolution in time of the solutions shown in panel (b).) 
}
\label{fig:purged}
\end{figure*}

For the truncated and purged equations, we
perform extensive direct numerical simulations, by using a standard
pseudo-spectral method and a $4^{\rm th}$ order Runge-Kutta scheme for
time-marching, on a $2\pi$-periodic line. We use two different sets of
collocation points, namely, $N = 16384$ and $N = 65536$ to obtain results for
$\kg = 500, 1000, 3000$ and 5000 (for $N = 16384$) and $\kg = 8000$, and 10000
(for $N = 65536$). For the purged simulations, additionally,
the theoretical estimates obtained, lead us to a choice of $\beta = 0.4,
0.6$ and $0.8$ and for each value of $\beta$, the inter-purging time was
obtained with $\alpha = 0.4, 0.6$, $0.8$, $0.9$, and $1.2$. (The simulations 
with $\alpha = 0.9$ and 1.2 were performed to confirm that too frequent purgings 
lead to thermalised solutions once more with the effective truncation wavenumber 
$K_p$.) 

The choice of time-steps in such simulations require some delicacy. For the
truncated problem, since the maximum principle is violated, individual
realisations of the velocity field can have excursions which are large (see
Fig.~\ref{thermalisation}b). Hence for the truncated simulations, as well as
those where purging is ineffective in preventing thermalisation, the time-step
$\delta t$ has to be kept very small. However, for the cases of
\textit{successful} purging (see manuscript), the maximum principle is no longer
violated. Hence for these cases we are able to choose $\delta t = 10^{-5}$ ($N
= 16384$) and $\delta t = 10^{-6}$ ($N = 65536$); for the analogous truncated
problem (and the ones where the $\alpha$-$\beta$ combination fail to prevent
thermalisation), $\delta t$ was taken to be at least two orders of magnitude
smaller.

In numerical simulations, $\delta t$ is typically set by the
resolution $\kg$ such that $\delta t \sim \mathcal{O}(\kg^{-1})$. As we have
noted before, purging if done too frequently would be akin to solving the
Galerkin-truncated Burgers equation with $\kg = K_p$.  This implies that
$\tau/\delta t \gg 1$ which, trivially, leads to $\alpha < 1$.  (We have
confirmed these conjectures through several, detailed numerical simulations.)

To obtain the entropy solution $u$, we use the Fast Legendre transform as discussed in
Refs.~\cite{Noullez} (see also Ref.~\cite{Mitra}) to solve the viscous Burgers equation 
in the vanishing viscosity $\nu \to 0$ limit. We solve the equation on a 2$\pi$ line 
with periodic boundary conditions and choose $N = 16384$ and $N = 65536$ collocation 
points (for easy comparison with the truncated and purged solutions; see below). The velocity 
field is evolved keeping in mind that the velocity potential $\psi$ (related to the 
velocity field via $u = -\partial_x\psi$) obeys a maximum principle:
\begin{equation}
\psi(x,t') = {\rm max}_y\left [ \psi(y,t) - \frac{(x-y)^2}{2(t'-t)}\right].
\end{equation}

Finally, we have studied the problem for several different initial conditions
(all of which consist of linear combinations of trigonometric polynomials
including the simplest single-mode case $\sin (x)$); we have checked that our
results and conclusions are consistent for all such initial conditions. In this
paper, for brevity, we present results only for the case $w_0 = v_0 = u_0 =
\sin (x) + \sin (2x+0.9) +\sin (3x)$.

In Fig.~\ref{fig:purged} we show representative plots, at $t
= 5.0$,  of the Galerkin-truncated $v$ (in blue and thermalised), the entropy $u$ (in black with a prominent shock)
and the purged solutions $w$ (in red) for (a) $\alpha = 0.6$, $\beta = 0.4$ and
(b) $\alpha = 0.8$, $\beta = 0.8$; we set the truncation wavenumber $\kg =
1000$. 
We immediately see that for $\alpha = 0.6$ and $\beta = 0.4$
(Fig~\ref{fig:purged}a), the solution $w$ approximates the entropy solution much
better---in so far as picking out the ramp structure and a jump near the
shock---though far from perfectly. 

Remarkably, if we choose $\alpha = \beta = 0.8$ (Fig~\ref{fig:purged}b)---and hence
much closer to satisfying the heuristic estimate $\alpha +\beta \lesssim 2$---the
agreement between the purged and entropy solutions are near-perfect.  Indeed
the main point of departure between the two solutions seems to be close to the
shock because of the ubiquitous Gibbs-type oscillations~\cite{NR}
associated with Fourier transforms of functions near discontinuities. 

We have checked that for $\alpha \gtrsim 0.9$, since $\tau/\delta t \sim \mathcal{O}(1)$,
the purged solutions thermalise once again as we conjectured. Hence, empirically, our 
extensive numerical simulations show that within the range of $\alpha$ that we study, the 
optimal choice is $\alpha = 0.8$. Furthermore, we have confirmed that our results are largely 
insensitive to the choice of $\beta$ as long as its greater than 1/3.

\begin{figure*}
\includegraphics[width=0.50\linewidth]{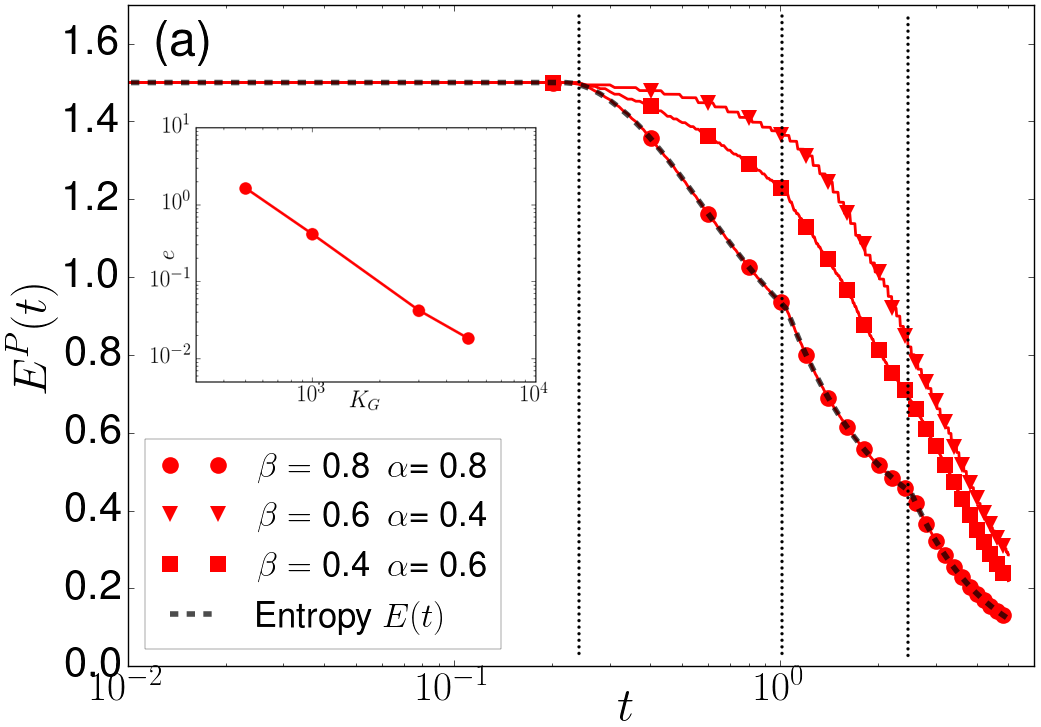}
\includegraphics[width=0.47\linewidth]{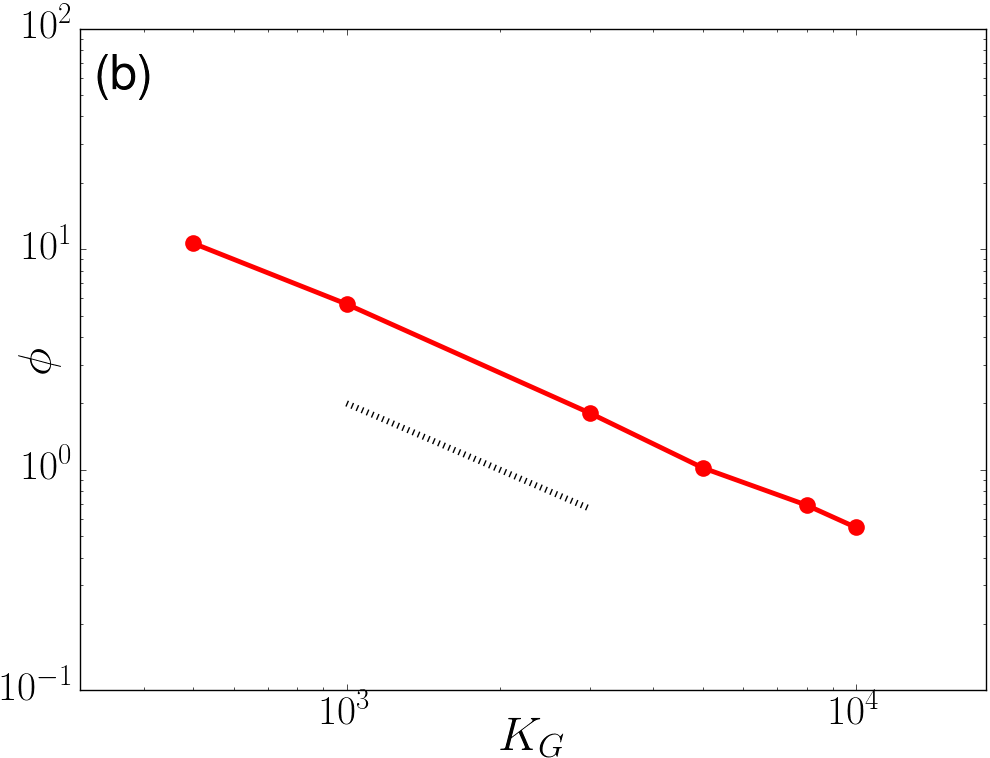}
\caption{(a) A plot of the total energy $E^{\rm P}$ versus time, from our purged solutions ~\eqref{purged}, for different 
combinations of $\alpha$ and $\beta$ and $\kg = 1000$. We also show, in black, the energy versus time plot for the entropy solution for comparison. 
The dashed vertical lines correspond to the times at which the shocks, three in all because of the three-mode initial conditions, form. In the inset, we 
plot the relative percentage error $e$ (see text) 
between the purged and entropy solution, for $\alpha = \beta = 0.8$, at $t = 5.0$, as a function of $\kg$. (b) A plot of the $L_2$ norm of 
the percentage relative error $\phi$ (see text) for $\alpha = \beta = 0.8$ as a function of $\kg$; the dashed-line shows 
a power-law $\kg^{-1}$ scaling consistent with the measured error.}
\label{error}
\end{figure*}

The fact that the purged and entropy solutions seem to be in agreement,
visually, suggests that the purged solution is dissipative as was anticipated,
by construction, earlier.  However, for this solution to actually converge to the entropy solution,
the rate of dissipation should be arbitrarily close to the dissipation
rate $\frac{dE}{dt}$ of the entropy solution. The most direct way to see this is to compare
the total energies of the entropy $E$ and the purged
$E^{\rm P}$ solutions, as a function of time, for different values of $\alpha$
and $\beta$: In Fig. ~\ref{error}(a) we show these results for $\kg = 1000$. We
find, as was already suggested in Fig.~\ref{fig:purged}, that for the optimal choice
$\alpha = \beta = 0.8$, the behaviour of the total energy versus time for the purged solution is
identical to the one obtained from the entropy solution. The purged solutions 
for other $\alpha-\beta$ combinations are dissipative as well; however 
they dissipate energy at rates much slower than the entropy solution.
Moreover, 
shock-mergers, as indicated by the vertical lines in the plot, and which lead
to tiny \textit{kinks} in the energy versus time profile, are faithfully
reproduced by purged solutions for $\alpha = \beta = 0.8$. 

A measure of how accurately the purged solution mimics the dissipation 
of the entropy one, is the percentage relative error 
$e = \frac{E^p - E}{E} \times 100$ at $t = 5.0$. In the inset of Fig.~\ref{error}(a), 
we plot $e$ as a function of $\kg$ for the most optimal purging choice 
($\alpha = \beta = 0.8$). Remarkably, this 
error $e$ decreases rapidly with $\kg$ and for $\kg = 5000$, $e \approx 0.01\%$. 

All of this leads us inevitably to the important question: For $\alpha = \beta
= 0.8$, does the purged solution indeed converge to the entropy one as $\kg \to
\infty$? A precise way to answer this is to measure the percentage
relative error (or the $L_2$ norm) $\phi = \sqrt{\frac{\displaystyle
\sum_{i=1}^{i = N} \left[ u(x_i)-w(x_i) \right]^2 }{\displaystyle \sum_{i=1}^{i
= N} \left[ u(x_i)\right]^2}} \times 100$ of the discrepancies
between the solutions $u$ and $w$. Given that this is a point-wise measure, unlike the global energy
measurements shown in Fig.~\ref{error}(a), a sharp decrease in $\phi$ with
$\kg$ should be clinching evidence of the efficacy of our scheme.  In
Fig.~\ref{error}(b), we show a log-log plot of $\phi$ as a function of $\kg$
and find a steep decrease ($\phi \sim \kg^{-1}$ indicated by the dashed line)
in the relative error as a function of $\kg$. For the large values of $\kg$,
the relative error $\phi < 1\%$, reaching a value of $\phi \approx 0.5\%$ for
$\kg = 10000$. 

These results show that purging leads to weak-dissipative solutions
which converge to the entropy solution of the parent PDE as $\kg \to \infty$.
Importantly, the discrepancy between the two solutions is already minute for
values of $\kg$ which are easily accessible. From the point of view of
numerical simulations, the $\beta \geq 1/3$ condition is extremely helpful
because it allows us to choose values of $\beta$ small enough such that for a
given $\kg$, the loss in resolution $\kg - K_p$ through purging, is
insignificantly small. As an example, for $\kg = 10000$ and $\beta = 0.4$,
fraction of resolution lost is about 0.3\%. 

\paragraph*{Summary and Outlook:} Our results, if seen in isolation for the Burgers equation, are admittedly
academic. This is because for the 1D Burgers equation, we have other ways to
obtain weak-dissipative solutions as well as the widths of the
analyticity strip $\delta$ analytically and numerically. Also, since for the
Burgers equation the effects of truncation are felt at times \textit{very}
close to $t_*$, the $\delta$ obtained for the Burgers
equation with and without purging, agree equally well with the theoretical estimate
up to times very close to $t_*$. This is pathological to the Burgers
equation and it is reasonable to conjecture that purging in the 3D Euler
equation will yield more dividends. Furthermore, there is no analogue of the Fast-Legendre method for  
the 3D Euler equations. 

It is in the light of the 3D Euler equations that this approach assumes special
importance. To the best of our knowledge, till date there is no algorithm which
allows, numerically, to obtain weak-dissipative solutions of the 3D Euler
equation. This algorithm allows us to do exactly that. Numerically, our
algorithm is trivial to implement in codes which solve the 3D
Galerkin-truncated Euler equation.  From earlier studies we know that the onset
of thermalisation in the 3D Galerkin-truncated Euler equation is formally
similar to that in the Burgers equation. Hence, the approach outlined in this
paper, should allow us to implement it for the 3D Euler equations and study,
numerically, dissipative solutions as well as, and possibly most importantly,
take advantage of the suppression of thermalisation to finally have a firm,
albeit numerical, answer for the celebrated blow-up problem. While 
it is true that for the 3D Euler equation, we are handicapped by a much poorer 
understanding of what the appropriate weak-dissipative solution ought to be, there are 
indeed several candidates against which our purged solutions may be benchmarked 
against, including the existing solutions of the incompressible Navier-Stokes equation 
for the largest Reynolds numbers currently attainable. We hope that our work will provide 
a stimulus for analogous (and important) studies of the truncated Euler equation.

Given the potential usefulness of our approach to revisit the analyticity strip method to 
numerically investigate the question of blow-up of the Euler equation, it might be useful to comment on 
recent studies of this problem. In brief, although there is some evidence that the Euler equations 
could avoid singularities through the formation of vortex sheets~\cite{Kerr_2013,Hussain1,Hussain2}, 
other results~\cite{Luo,Elgindi, Mailybaev} suggests that this question is far from settled. Therefore, 
our work, although demonstrated here for the Burgers equation, could play a role in revisiting 
this issue from the point of view of the width of the analyticity strip. In this context, it may be 
worth recalling that the one of the earliest demonstrations of the analyticity strip method for the 
Galerkin-truncated inviscid hydrodynamics, was for the Burgers equation~\cite{Sulem1983}.

Before we conclude, it is important to ask if thermalisation can be suppressed
by other means (without resorting to viscosity).  Purging attempts in physical
space---which consists in {\it smoothening} out the \textit{tygers} in physical
space through local averaging---does not result in any significant suppression
of thermalisation and lacks easy adaptability to different initial conditions
and higher dimensional equations.  A second possibility is of course the use of
a hyperviscous term. This however has the drawback that we would end up solving
not the inviscid equation but its viscous form and for higher-orders of the
hyperviscosity---which is similar in spirit to the idea of purging---the
solutions thermalise~\cite{Frisch-PRL-2008,Banerjee-PRE-2014,Agrawal}.  Another
approach is due to Pereira, {\it et al.}~\cite{PereiraPRE} who showed that a
wavelet-based filtering technique also leads to a suppression of the resonances
leading to \textit{tygers}. However, such an approach has the limitation, as
mentioned by the authors themselves, that the dual operations of filtering and
truncation at every time step do not commute. Hence the weak dissipation
introduced in this approach is somewhat uncontrolled.  To this extent we feel
that the prescription we present here is most suited for generating weak-dissipative
solutions and, importantly, more easily adaptable to higher-dimensional systems
such as the 3D Euler equations.  

\textit{Note added:} We recently became aware of the work by Fehn \textit{et al.} on 
obtaining evidence for the anomalous energy dissipation in the Euler equation~\cite{fehn2020numerical} 
by using a higher order discontinuous Galerkin discretization developed by the authors. These 
results show promise for a numerical resolution of Onsager's conjecture by using methods 
different from ours.

\begin{acknowledgments}
SSR and UF acknowledge the support of the Indo-French Centre for
Applied Mathematics. SDM and SSR  acknowledge support of the DAE,
Govt.~of~India, under project no.~12-R\&D-TFR-5.10-1100.
SSR acknowledges DST (India) project MTR/2019/001553 for support. 
UF, NB and SN acknowledge the support of Universit\'e C\^ote d'Azur. 
\end{acknowledgments}
\bibliography{references} 
\end{document}